\documentclass[twocolumn,prl,aps,twocolumn]{revtex4-1}
\usepackage[latin9]{inputenc}
\usepackage{amsmath}
\usepackage{amssymb}
\usepackage{graphicx}
\usepackage[unicode=true,pdfusetitle,
 bookmarks=false,
 breaklinks=false,pdfborder={0 0 1},backref=false,colorlinks=false]
 {hyperref}
\hypersetup{
 bookmarksnumbered=false,bookmarksopen=false}

\makeatletter
\@ifundefined{textcolor}{}
{%
 \definecolor{BLACK}{gray}{0}
 \definecolor{WHITE}{gray}{1}
 \definecolor{RED}{rgb}{1,0,0}
 \definecolor{GREEN}{rgb}{0,1,0}
 \definecolor{BLUE}{rgb}{0,0,1}
 \definecolor{CYAN}{cmyk}{1,0,0,0}
 \definecolor{MAGENTA}{cmyk}{0,1,0,0}
 \definecolor{YELLOW}{cmyk}{0,0,1,0}
}

\usepackage{color}

\@ifundefined{textcolor}{}{%
 \definecolor{BLACK}{gray}{0} 
 \definecolor{WHITE}{gray}{1}
 \definecolor{RED}{rgb}{1,0,0}
 \definecolor{GREEN}{rgb}{0,1,0}
 \definecolor{BLUE}{rgb}{0,0,1}
 \definecolor{CYAN}{cmyk}{1,0,0,0}
 \definecolor{MAGENTA}{cmyk}{0,1,0,0}
 \definecolor{YELLOW}{cmyk}{0,0,1,0}
}

\usepackage{txfonts}

\makeatother

\begin{document}

\title{Soft Quantum Control for Highly Selective Interactions among Joint Quantum Systems}

\author{J. F. Haase}
\affiliation{Institut f\"ur Theoretische Physik und IQST, Albert-Einstein-Allee
11, Universit\"at Ulm, D-89069 Ulm, Germany}

\author{Z.-Y. Wang}
\email{zhenyu3cn@gmail.com}
\affiliation{Institut f\"ur Theoretische Physik und IQST, Albert-Einstein-Allee
11, Universit\"at Ulm, D-89069 Ulm, Germany}

\author{J. Casanova}
\affiliation{Institut f\"ur Theoretische Physik und IQST, Albert-Einstein-Allee
11, Universit\"at Ulm, D-89069 Ulm, Germany}

\author{M. B. Plenio}
\email{martin.plenio@uni-ulm.de}
\affiliation{Institut f\"ur Theoretische Physik und IQST, Albert-Einstein-Allee
11, Universit\"at Ulm, D-89069 Ulm, Germany}

\begin{abstract}
We propose a quantum control scheme aimed at interacting systems that gives rise to highly selective coupling among their near-to-resonance constituents.  
Our protocol implements temporal control of the interaction strength, switching it on and off again adiabatically. 
This soft temporal modulation significantly suppresses off-resonant contributions in the interactions.
Among the applications of our method we show that it allows us to perform an efficient rotating-wave approximation in a wide parameter regime, the elimination of side peaks in quantum sensing experiments, and selective high-fidelity entanglement gates on nuclear spins with close frequencies.  We apply our theory to nitrogen-vacancy centers in diamond and demonstrate the possibility for the detection of weak electron-nuclear coupling under the presence of strong perturbations.
\end{abstract}

\maketitle
\emph{Introduction.--} The ability to selectively manipulate and couple the constituents in an interacting quantum cluster is a fundamental
requirement for a wide range of technological applications~\cite{Heremans16,Prati13,Rong17}.
For instance, the {\it individual addressing} of magnetic nuclei in a target  molecule with a quantum sensor, such as the nitrogen-vacancy (NV) center in diamond~\cite{Doherty13}, is a crucial requirement to determine the 3D structure of single molecules of interest for biochemistry and medicine~\cite{Schirhagl14, Wu16, Zhao11Atomic,Shi14Sensing,Wang16,Ma16Angstrom,Boss16,Ma16Proposal}. In addition, the selective coupling of the quantum sensor with nearby quantum registers would enhance the sensitivity and resolution of quantum sensing protocols~\cite{Unden16,Lovchinsky16,Zaiser16,Wang17,Matsuzaki,Rosskopf16}.
From a different point of view, if the addressing operation does not disturb the other qubits surrounding a certain target register, namely the $^{13}$C
and $^{29}$Si nuclear spins that appear in diamond~\cite{Liu2013Noise,Taminiau14,Waldherr2014Quantum,Mueller14,Mkhitaryan15}
and silicon carbide~\cite{Baranov05,Seo16}, or Eu$^{3+}$ ions in stoichiometric rare-earth crystals~\cite{Ahlefeldt10,Ahlefeldt16}, one can use the available qubits for quantum information~\cite{Casanova16,Perlin:1708.09414,Abobeih:1801.01196}
or quantum simulation~\cite{Cai13Large} tasks. Furthermore, nuclear qubits
coupled to an electron spin are also important to build a robust optical interface
for quantum networks~\cite{Reiserer16,Kalb17}. 

\begin{figure}[t!]
\includegraphics[width=0.95\columnwidth]{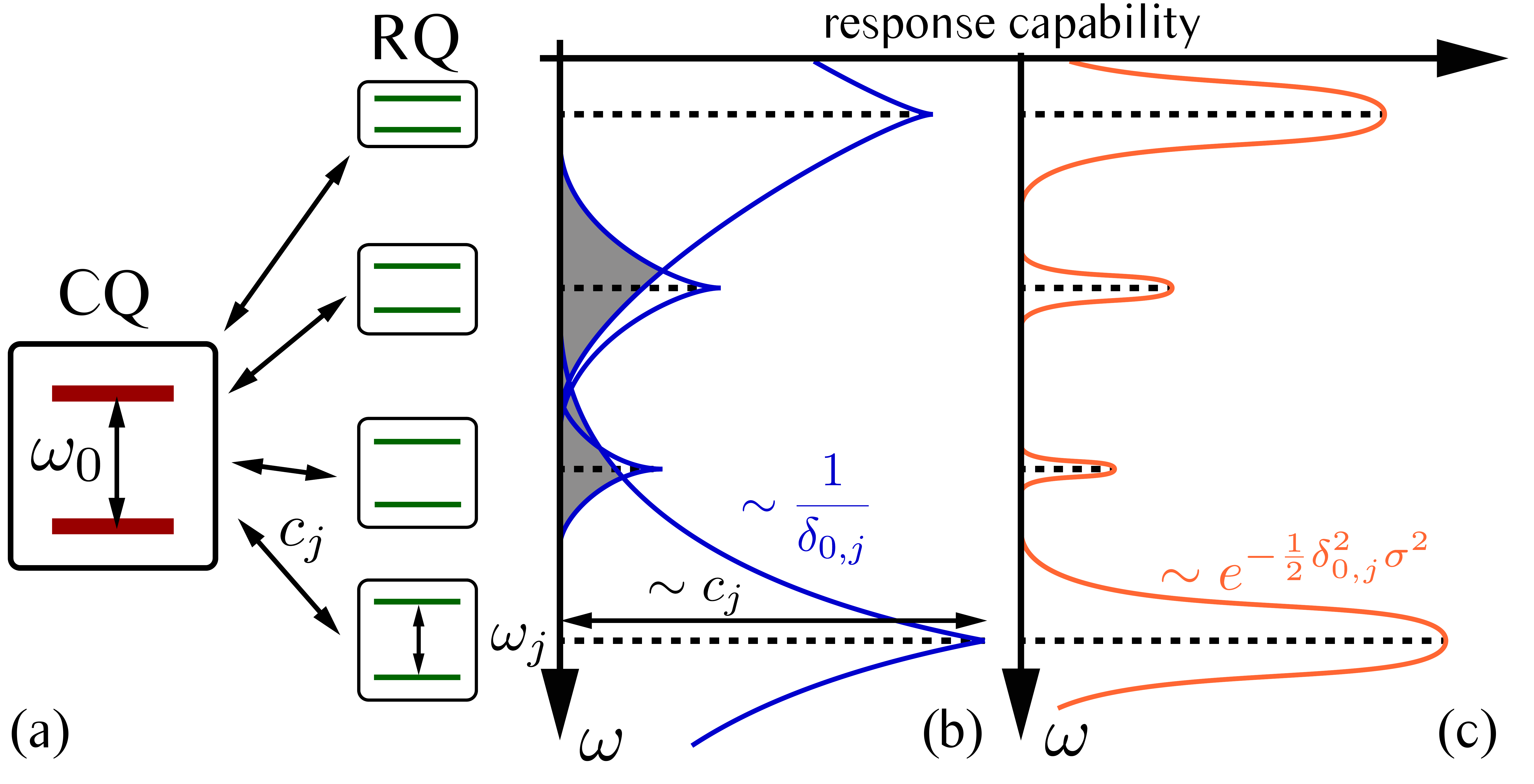} \caption{Advantages of temporal shaping of coupling. (a) Illustration
for the control qubit and resource qubits. (b) For the case of constant
coupling $c_{j}$, the off-resonant response
decays slowly $c_{j}/\delta_{0,j}$ (blue lines) with the energy mismatch
$\delta_{0,j}$. The overlaps (gray areas) on the frequency response
prohibit high-fidelity selective coupling. (c) With the soft coupling
proposed in this work the off-resonant response decays exponentially
(see the orange lines), which allows high-fidelity
addressing. }
\label{fig:obstacles} 
\end{figure}

The addressability problem can be reduced to the situation shown in Fig.~\ref{fig:obstacles} (a)
where a control qubit (CQ) interacts with multiple resource qubits
(RQs)~\cite{Chen17,Casanova17Arbitrary}. In order to exert control on a certain RQ the characteristic frequency $\omega_{0}$ of the CQ
is tuned to the resonance frequency $\omega_j$ of the RQ via a 
continuous drive that exploits the Hartmann-Hahn resonance
\cite{Hartmann62,Cai13,London13Detecting} or the application of pulsed dynamical decoupling (DD)~\cite{Yang11,Souza12,Kolkowitz2012Sensing,Taminiau2012Detection,Zhao2012Sensing}.
As we will show later,
because of the time independent coupling $c_{j}$ between the CQ and each RQ, the spectral
responses are proportional to $c_{j}/\delta_{0,j}$ which decays slowly with the energy mismatch
$\delta_{0,j}=\omega_{0}-\omega_{j}$ ($j>0$ for RQs), i.e. in a power-law manner. Therefore other off-resonant RQs will considerably perturb
the CQ and vice versa, see Fig.~\ref{fig:obstacles} (b), prohibiting
the high-fidelity addressing on the desired target RQ. 
This is particularly challenging for realistic settings where the RQs only slightly differ in their resonance frequencies.

In this Letter we propose the idea of soft temporal quantum control which enables on-resonant coupling within a desired set of target systems, while efficiently avoiding unwanted off-resonant contributions coming from others. With the specific case of Gaussian soft control,  off-resonant
effects are exponentially suppressed by the mismatch $\delta_{0,j}$ as $\exp(-\sigma^{2}\delta_{0,j}^{2}/2)$, see Fig.~\ref{fig:obstacles} (c), achieving high-selective coupling. In addition, we develop an average Hamiltonian theory for our soft quantum control method. By using the quantum adiabatic theorem, we take high-order virtual transitions into account and provide an accurate description of the dynamics even for situations involving strong perturbations and long evolution
times. We will show specific applications of our method such as the realization of an efficient rotating-wave approximation (RWA)
and highly selective two-qubit gates for quantum sensing and computing.

\emph{Generic model.--} To explore the
effects emerging from temporal control,
we consider the Hamiltonian ($\hbar=1$) $H=H_{{\rm S}}+H_{{\rm int}}$, where $H_{{\rm S}}$ is the Hamiltonian for the
quantum registers and 
$H_{{\rm int}}=\lambda(t)\sum_{\alpha}c_{\alpha}V_{\alpha}$
describes their interactions that we want to perform selective control.
$\lambda(t)$ is a dimensionless, time-dependent global factor and 
$V_{\alpha}$ may be single-body or $N$-body operators with strength $c_{\alpha}$ (i.e. the norm of $V_{\alpha}$ is bounded to one).

In terms of the eigenvalues $\omega_{j}$ and the projection operators $\mathbb{P}(\omega_{j})$ of the Hamiltonian $H_{{\rm S}}=\sum_{j}\omega_{j}\mathbb{P}(\omega_{j})$,
we write
\begin{equation}
H_{{\rm int}}=\lambda(t)\sum_{\alpha,j,k}c_{\alpha}V_{\alpha}^{\omega_{j},\omega_{k}},\label{eq:interactionDecomposition}
\end{equation}
where each $V_{\alpha}^{\omega_{j},\omega_{k}}\equiv\mathbb{P}(\omega_{j})V_{\alpha}\mathbb{P}(\omega_{k})$ fulfils
\begin{equation}
[H_{{\rm S}},V_{\alpha}^{\omega_{j},\omega_{k}}]=(\omega_{j}-\omega_{k})V_{\alpha}^{\omega_{j},\omega_{k}}.\label{eq:HsVCommutator}
\end{equation}
In coupled quantum networks, $V_{\alpha}^{\omega_{j},\omega_{k}}$
would describe the interaction between quantum systems with an energy
mismatch of $\delta_{j,k}\equiv\omega_{j}-\omega_{k}$. 
Our target is to suppress the terms $V_{\alpha}^{\omega_{j},\omega_{k}}$ in Eq.~\eqref{eq:interactionDecomposition} for which $\omega_{j}\neq\omega_{k}$, and to keep the energy conserving ones (i.e. those with $\delta_{j,k}=0$) by shaping the parameter $\lambda(t)$ for the sake of enhanced selectivity.

\emph{Leading-order effects and soft quantum control.--} In a rotating frame with respect to $H_{{\rm S}}$, $H_{{\rm int}}$ becomes
$H_{{\rm int}}^{\prime}(t)=\lambda(t)\sum_{\alpha,j,k}c_{\alpha}V_{\alpha}^{\omega_{j},\omega_{k}}e^{i\delta_{j,k} t}$.
In the absence of a modulation for $\lambda$, i.e. $\lambda(t)=\lambda_{0}$,
unwanted terms in $V_{\alpha}$ can be neglected by the RWA provided
that the $\delta_{j,k}$ is sufficiently large compared with $\lambda_{0}c_{\alpha}$. To see how the modulation of $\lambda(t)$  improves this situation,
we calculate the leading-order
effective Hamiltonian in the rotating frame by using the Magnus expansion~\cite{Haeberlen68,Mananga16} 
for a time interval $[-T/2,T/2]$; it reads
\begin{equation}
\bar{H}_{{\rm int}}^{(1)}=\frac{1}{T}\int_{-T/2}^{T/2}dtH_{{\rm int}}^{\prime}(t)=\sum_{\alpha,j,k}c_{\alpha}g(\delta_{j,k})V_{\alpha}^{\omega_{j},\omega_{k}},\label{eq:HBarInt}
\end{equation}
where the averaging factor 
\begin{equation}
g(\delta_{j,k})=\frac{1}{T}\int_{-T/2}^{T/2}dt\lambda(t)e^{i\delta_{j,k} t}\label{eq:gDelta}
\end{equation}
can be controlled by $\lambda(t)$.

For the conventional case of a constant $\lambda(t)=\lambda_{0}$, we have 
\begin{equation}
g(\delta_{j,k})=g_{{\rm C}}(\delta_{j,k})\equiv\lambda_{0}\frac{\sin\left(T\delta_{j,k}/2\right)}{\left(T\delta_{j,k}/2\right)}.\label{eq:gC}
\end{equation}
In this manner, unwanted terms in $V_{\alpha}$ are suppressed by
a large energy mismatch $\delta_{j,k}$
to decrease the value of $g_{{\rm C}}(\delta_{j,k})$~\cite{Note:Detuning}.
By selecting $\lambda_{0}$ sufficiently small, the off-resonant interactions can be more efficiently suppressed
with the associated improvement in the addressing for the resonant terms.  See Refs.~\cite{Wang16,Casanova15}
for a specific application of the latter to
the case of NV centers in diamond surrounded by $^{13}$C nuclear
spins. However, from Eqs.~(\ref{eq:HBarInt}) and (\ref{eq:gC}) the effects introduced by off-resonant
terms decay slowly as a power law $\lambda_{0}c_{\alpha}/\delta_{j,k}$
on the energy mismatch $\delta_{j,k}$ while, in addition, because $g(0)=\lambda_{0}$ a decrease on $\lambda_{0}$
also carries the undesired effect of reducing the intensity of the coupling with the resonant terms. 

From Eq.~(\ref{eq:gDelta}), we find that by using a time-dependent
soft modulation, i.e., $\lambda(t)$ is small at the beginning
and at the end of quantum evolution, the non-resonant terms can be removed
with greater fidelity. More specifically, we propose the Gaussian temporal modulation
\begin{equation}
\lambda(t)=\lambda_{0}\exp\left[-t^{2}/(2\sigma^{2})\right],\label{eq:lambdaGaussian}
\end{equation}
which has the corresponding factor
\begin{equation}
g(\delta_{j,k})=g_{{\rm M}}(\delta_{j,k})\equiv\lambda_{0}\eta(\sigma,T)\exp\left(-\frac{1}{2}\sigma^{2}\delta_{j,k}^{2}\right),\label{eq:GaussCouplingMod}
\end{equation}
where $\eta(\sigma,T)=\sqrt{\frac{\pi}{2}}\frac{\sigma}{T}\left[\text{erf}\left(\frac{T-2i\sigma^{2}\delta_{j,k}}{2\sqrt{2}\sigma}\right)+\text{erf}\left(\frac{T+2i\sigma^{2}\delta_{j,k}}{2\sqrt{2}\sigma}\right)\right]$
and $\text{erf}(x)=\frac{2}{\sqrt{\pi}}\int_{0}^{x}dze^{-z^{2}}$.
A simple inspection of Eq.~(\ref{eq:GaussCouplingMod}) reveals that the effective couplings $g_{{\rm M}}(\delta_{j,k})c_{\alpha}$ decay exponentially
with  $\delta_{j,k}$. Hence, we expect the selectivity to be dramatically improved. We want to remark that our temporal shaping
scheme shares interesting similarities with the control by Gaussian pulses
of classical fields~\cite{Vandersypen05}, however, in our case, the
shaping is exerted on the coupling between quantum systems where quantum
backaction plays a significant role on both sides~\cite{Zhao11Anomalous}.

\emph{Higher-order effects and adiabatic average Hamiltonian.--} Although
the leading-order average Hamiltonian $\bar{H}_{{\rm int}}^{(1)}$ in Eq.~\eqref{eq:HBarInt}
describes well the dynamics for $T\ll1/\text{max}|c_{\alpha}|$, if
strong coupling constants are present, higher-order corrections~\cite{Haeberlen68,Mananga16} have
to be included in order to have an accurate description of the dynamics for 
larger times. 

While the evaluation of higher-order terms is involved in the general case, now we will show that our proposed soft quantum control scheme allows us to easily describe the system propagator including
high-order corrections when executed in an adiabatic manner. To this end we first analyze the propagator
$U_{{\rm D}}=\exp\left(-i\int_{-T/2}^{T/2}H_{{\rm D}}dt\right)$,
where $H_{{\rm D}}=H_{{\rm S}}+\lambda(t)\sum_{\alpha}c_{\alpha}\sum_{j}V_{\alpha}^{\omega_{j},\omega_{j}}$
includes the on-resonance desired interactions. In the latter all $V_{\alpha}^{\omega_{j},\omega_{j}}$ operators
commute with $H_{{\rm S}}$, see Eq.~(\ref{eq:HsVCommutator}), hence
$H_{{\rm D}}$ can be diagonalized in the common eigenstates $|\psi_{n}^{{\rm D}}\rangle$
($n=1,2,\ldots$) of $H_{{\rm S}}$ and $V_{\alpha}^{\omega_{j},\omega_{j}}$.
Therefore $U_{{\rm D}}=\sum_{n}e^{-i\phi_{n}^{{\rm D}}(T)}|\psi_{n}^{{\rm D}}\rangle\langle\psi_{n}^{{\rm D}}|$
is also diagonal in the basis $\{|\psi_{n}^{{\rm D}}\rangle\}$ and
the dynamic phases $\phi_{n}^{{\rm D}}(T)$ include the effect of
energy shifts coming from $V_{\alpha}^{\omega_{j},\omega_{j}}$.

If the whole Hamiltonian $H$ is considered, the time-ordered evolution
$U=\mathcal{T}\exp\left[-i\int_{-T/2}^{T/2}H(t)dt\right]$ is generally
non-diagonal in the basis $\{|\psi_{n}^{{\rm D}}\rangle\}$ and the non-commuting $V_{\alpha}^{\omega_{j},\omega_{k}}$ terms 
would cause unwanted transitions between the different $|\psi_{n}^{{\rm D}}\rangle$ states.

However, when the soft control is included one can efficiently eliminate the unwanted
interactions caused by $V_{\alpha}^{\omega_{j},\omega_{k}}$,
even for long evolution times $T$. 
At the boundaries of the interaction
times ($-T/2$ and $T/2$), $\lambda(t)$ has negligible values and
therefore the system's eigenstates coincide with those of $H_{{\rm D}}$.
More precisely, under the condition of adiabatic evolution~\cite{Wang16Necessary,Xu:1711.02911},
there are no transitions among the states $|\psi_{n}^{{\rm D}}\rangle$
and the propagator at the end of the evolution is $U\approx\sum_{n}e^{-i\phi_{n}(T)}|\psi_{n}^{{\rm D}}\rangle\langle\psi_{n}^{{\rm D}}|\equiv\bar{U}\equiv e^{-i \bar{H}T}$,
where $\phi_{n}(T)$ are the dynamic phases, while the geometric phases vanish because $\lambda(t)$ returns to its original value \cite{Griffiths05Introduction}. In
this manner $U$ takes the same form as $U_{{\rm D}}$ and the adiabatic average Hamiltonian for the soft quantum control scheme
\begin{equation}\label{aaH}
\bar{H}=\sum_{n}\left[\phi_{n}(T)/T\right]|\psi_{n}^{{\rm D}}\rangle\langle\psi_{n}^{{\rm D}}|,
\end{equation}
is diagonal in the same basis as $H_{{\rm D}}$ and includes all the high-order energy shifts.
In the following we illustrate our general theory via 
two important applications.


\begin{figure}[t!]
\includegraphics[width=0.95\columnwidth]{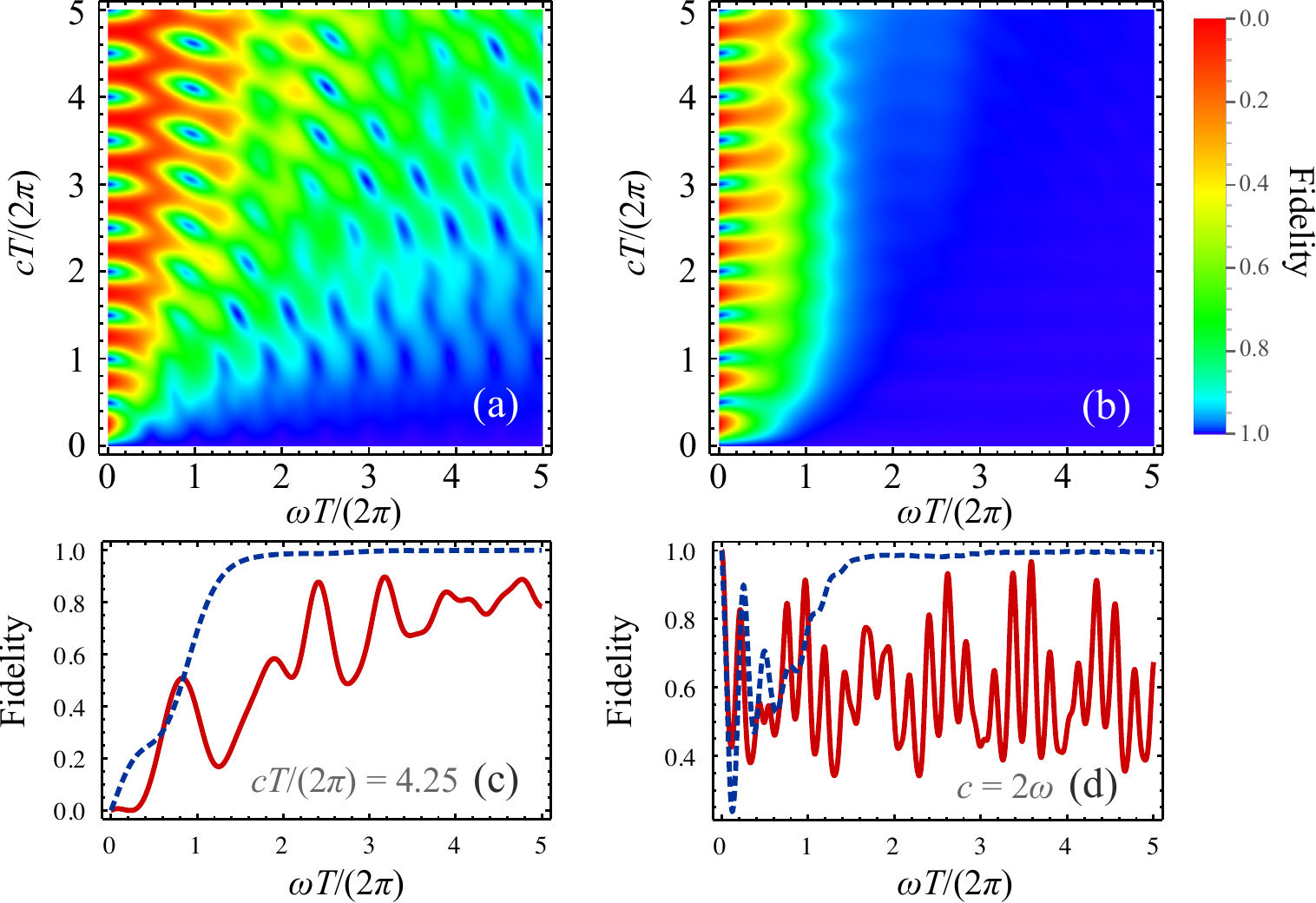} \caption{(Color online) Fidelities under RWA. (a) Fidelity to the target evolution
without unwanted coupling by using the constant-amplitude coupling.
(b) As in (a) but using a Gaussian soft coupling with $\sigma=T/(4\sqrt{2})$
and $1/\lambda_{0}=\sqrt{2\pi}\sigma\text{erf}\left(\frac{T}{2\sqrt{2}\sigma}\right)$
such that we obtain the same target evolution. Curves in (c) and (d) show 
cross-sectional plots in the constant-amplitude case of (a) [red solid
lines], or for the Gaussian shaped coupling case of (b) [blue dashed lines].
It is easy to see that the soft quantum control scheme keeps a high fidelity even for a relatively large ratios $c/\omega$ at long evolution times $T$.}
\label{fig:RWA} 
\end{figure}

\emph{Improved RWA.--} Here we demonstrate how
the soft quantum control mechanism efficiently eliminates the non-energy-conserving (or counter rotating) 
terms over a continuous time interval even for long evolution times. 
The existence of the non-energy-conserving terms is due to the limit of available resources for selective control on realistic quantum systems (e.g., singlet-triplet qubits in semiconductor quantum dots~\cite{Wardrop14,Nichol2017}). As an example,
we consider a control qubit ($0$) and two equally strong coupled resource qubits ($1,\,2$) (e.g., singlet-triplet qubits~\cite{Wardrop14,Nichol2017})
with the interaction
\begin{equation}
H_{{\rm int}}=c\lambda(t)\sigma_{0}^{x}\left(\sigma_{1}^{x}+\sigma_{2}^{x}\right)=c\lambda(t)\left(P_{0,1}+Q_{0,1}+\sigma_{0}^{x}\sigma_{2}^{x}\right),
\end{equation}
where $\sigma_{j}^{\alpha}$ ($\alpha=x,y,z$) denotes a Pauli operator for the $j$-th qubit,
$P_{0,1}=\sigma_{0}^{+}\sigma_{1}^{-}+{\rm H.c.}$ with $2\sigma_{j}^{\pm}=\sigma_{j}^{x}\pm i\sigma_{j}^{y}$, and $Q_{0,1}=\sigma_{0}^{+}\sigma_{1}^{+}+\sigma_{0}^{-}\sigma_{1}^{-}$.
The coexistence of $P_{0,1}$ and $Q_{0,1}$ can be due to the nature of systems (see \cite{Nichol2017} for a realistic example).
We aim to interact qubit 0 purely with qubit 1 via the flip-flop term $P_{0,1}$ without involving the perturbation $Q_{0,1}$. 
Therefore the energies $H_{{\rm S}}=\frac{\omega}{2}\left(\sigma_{0}^{z}+\sigma_{1}^{z}\right)+\frac{\omega_2}{2}\sigma_2^z$ are chosen such that qubit 2 is off resonant with $\omega_2 = 3\omega$.
The corresponding target Hamilton
\begin{equation}
H_{{\rm target}}=\frac{1}{2}\tilde{\omega}\left(\sigma_{0}^{z}+\sigma_{1}^{z}\right)+\frac{1}{2}\tilde{\omega}_2\sigma_2^z+\tilde{c}\left(\sigma_{0}^{+}\sigma_{1}^{-}+{\rm H.c.}\right),
\end{equation}
with the associated propagator $U_{{\rm target}}=e^{-iH_{{\rm target}}T}$ can be used to generate high-fidelity swap gate
between qubits 0 and 1. 
The corrected energies and interaction marked with a tilde can be obtained by using the adiabatic average Hamiltonian according to Eq.~(\ref{aaH}).

In Fig.~\ref{fig:RWA} (a) and with the red solid lines in (c) and
(d), we show the gate fidelities $F=|{\rm Tr}(U_{{\rm target}}U^{\dagger})|/{\rm Tr}(UU^{\dagger})$~\cite{WangGateFid} (here $U=\mathcal{T}e^{-i\int_{-T/2}^{T/2}(H_{{\rm S}}+H_{{\rm int}})dt}$)
with respect to the target evolution $U_{{\rm target}}$ for the standard coupling $\lambda(t)=1$,
while in Fig.~\ref{fig:RWA} (b) and with the blue dashed lines in
(c) and (d) the fidelities are plotted for a situation involving the
soft quantum modulation in Eq.~(\ref{eq:lambdaGaussian}). An
inspection of these plots reveals that the soft coupling approach results in
much higher fidelities in a wide range of parameters, even for
strong coupling regimes $(c>\omega)$ and 
a wide range of evolution times. 
In contrast, the standard approach does not achieve a high fidelity to the target Hamiltonian because an efficient elimination 
of the oscillating terms requires weak couplings and longer averaging periods. Naturally during these times,
relaxation and decoherence processes will decrease the fidelity further. Furthermore, locating the points 
of high fidelity in the standard approach becomes increasingly difficult when more qubits are involved (cf. the two-qubit example in the Supplemental Material~\cite{SuppMat}).  

Note that our approach is fundamentally different from adiabatic elimination~\citep{Torosov12, Vitanov17}. 
Adiabatic elimination is aimed at \emph{coupling} certain target levels by a virtual transfer of excitations through other mediator states that are removed from the dynamics, thus generating an evolution in the \emph{reduced} Hilbert space of the target states. Instead our objective is to efficiently \emph{suppress} unwanted interaction terms in the Hamiltonian through a soft modulation of the coupling constants, without reducing the dimension of the whole Hamiltonian and without having to use other states as mediators. Hence our method allows us to switch off unwanted interactions among the qubits in a highly selective manner and to perform high-fidelity quantum gates as we will demonstrate later.

\begin{figure}[t!]
\begin{centering}
\vspace{0cm}
 \hspace{-0.3cm} \includegraphics[width=0.95\columnwidth]{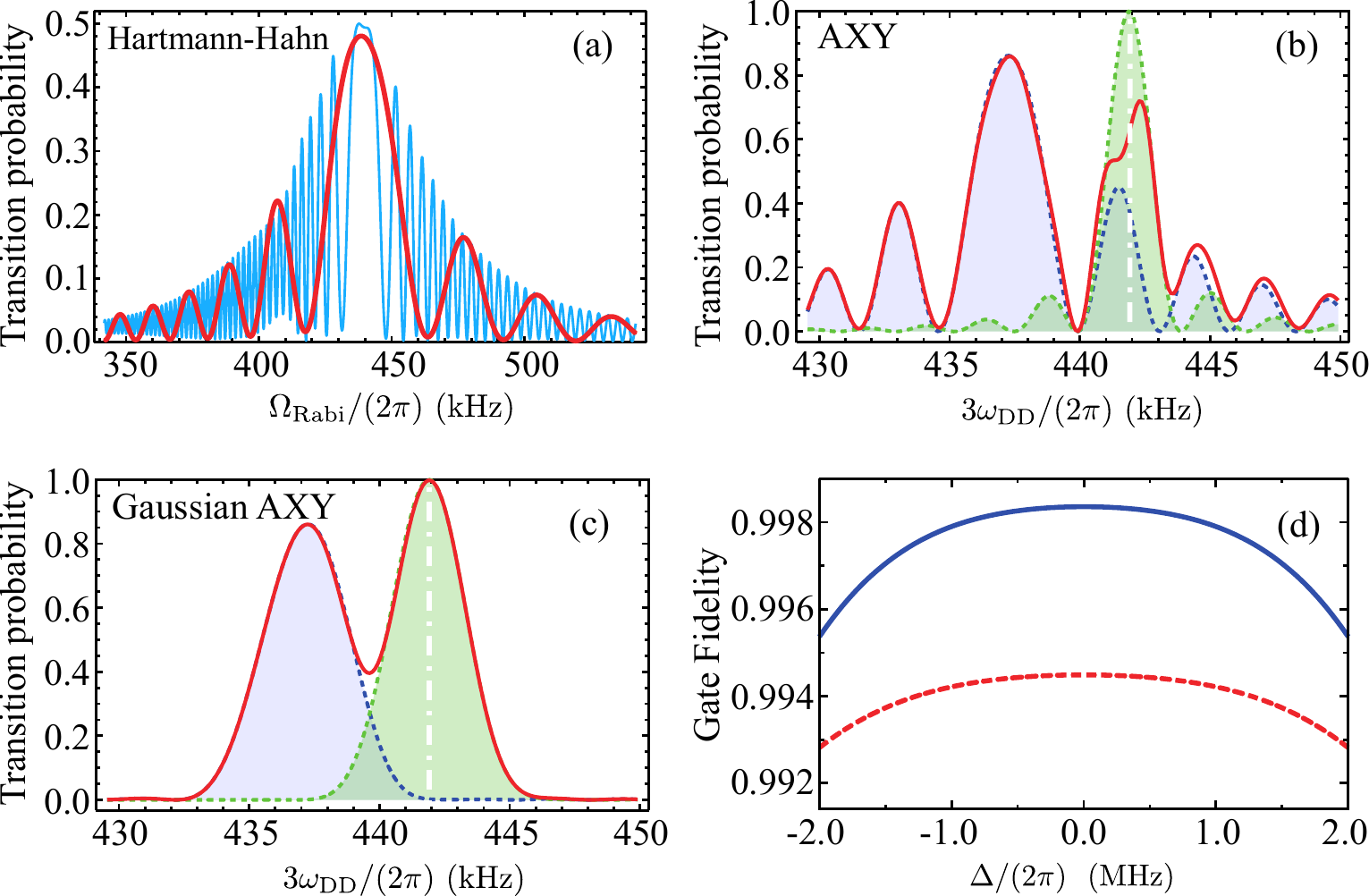} 
\par\end{centering}
\caption{(Color online) (a)-(c) Signal of transition probabilities, originating from two nuclear spins 
($\omega_{1}=2\pi\times441.91$ kHz and $\omega_{2}=2\pi\times437.54$ kHz, see \cite{SuppMat} for more details).
(a) Hartmann-Hahn resonance spectrum with a
total sensing time $T\approx54$ $\mu$s (red) or $T\approx435$
$\mu$s (blue). (b) Signal for AXY sequences (red solid line) at the
third harmonic with 128 composite pulses each has five elementary $\pi$ pulses). 
The single-spin contributions are drawn with dashed lines
and corresponding shading, and $f_{3}=0.271$ is chosen to maximize the one from the first spin.
The target signal centered at the vertical
dash-dotted line is destroyed by the strong perturbation from the
unwanted second spin. (c) Varying $f_{3}$ of the AXY
sequences in (b) according to the Gaussian shape
clearly resolves the two spins.
(d) The fidelity (blue solid curve) of the gate 
$U_\text{target}=e^{-i\frac{\pi}{4}\sigma_{0}^{z}\sigma_{1}^{x}}\otimes \mathbb{I}_{2}$
as a function of the microwave detuning error $\Delta$ by using the Gaussian AXY sequence with
a Rabi frequency $\Omega=2\pi\times 20$~MHz in the rectangular pulses. The red dashed line
is the case for a mismatch of $5\%$ in $\Omega$.
To realize the gate $U_\text{target}$, $f_{3}$ has been reduced by a factor of two when using the parameters indicated by the vertical
dashed-dotted line in (c).}\label{fig:application} 
\end{figure}

\emph{Selective qubit addressing.--} The soft quantum control mechanism allows
high-fidelity interactions between weakly coupled qubits while it avoids
perturbations that arise from the presence of strongly coupled qubits. Since the NV center in
diamond is an excellent platform for quantum information processing~\cite{Casanova16,Perlin:1708.09414,Abobeih:1801.01196}, quantum networks~\cite{Reiserer16,Kalb17}
and quantum sensing~\cite{Schirhagl14, Wu16}, we consider a network consisting of an NV electron spin
and its surrounding $^{13}{\rm C}$ nuclear spins (see~\cite{SuppMat} for details of the model). 

The electron-nuclear hyperfine coupling offers a medium to control the $^{13}{\rm C}$ nuclear spins via the NV electron.
Under pulsed DD~\cite{Kolkowitz2012Sensing,Taminiau2012Detection,Zhao2012Sensing,Casanova15} or a continuous drive~\cite{Cai13,London13Detecting} on the NV electron states $m_{\rm{s}}=0$ and, say, $m_{\rm{s}}=-1$, the $^{13}{\rm C}$ Larmor frequencies $\omega_j$ are shifted by hyperfine coupling, 
providing the frequency differences $\delta_{j,n}=\omega_{j}-\omega_{n}$ for selective addressing~\cite{SuppMat}. However, 
the differences $\delta_{j,n}$ and the electron-nuclear interactions are typically of the same order of magnitude, imposing a challenge on 
highly selective coupling.

To demonstrate the advantages of the soft quantum control, we compare different protocols in Fig.~\ref{fig:application} 
by using a model with two spectrally close nuclear spins (their coupling is small but is taken into account in simulations). 
As shown in Fig.~\ref{fig:application}~(a) 
a frequency scan obtained via a continuous, constant drive~\cite{Cai13,London13Detecting} does not resolve the two $^{13}{\rm C}$ nuclei
even for a longer sensing time $T$ because of the slow power-law-decay of the signal around
the resonance position. 

Because pulsed DD sequences can be implemented easily in current experimental setups and
coherent control on NV electron and nuclear spins longer than one second has been 
experimentally implemented with over ten thousands of DD pulses~\cite{Abobeih:1801.01196},
we apply a DD sequence to preserve the coherence of the NV
electron qubit and to realize the Hamiltonian~\cite{SuppMat}
\begin{equation}
H=-\frac{1}{8}f_{k_{{\rm DD}}}\sigma_{0}^{z}\sum_{j>0}a_{j}^{\perp}\sigma_{j}^{x}-\sum_{j>0}\frac{1}{2}\delta_{j,n}\sigma_{j}^{z},
\end{equation}
for addressing the nuclear frequency $\omega_{n}$. Both $\delta_{j,n}$ and $a_{j}^{\perp}$
are determined by the hyperfine coupling at the nuclear locations.

The DD protocol of adaptive-XY (AXY) sequences~\cite{Casanova15} provides better performance~\cite{Casanova15,Wang16} over standard
DD sequences~\cite{Yang11,Souza12} because it has strong robustness against control errors
and allows us to tune $f_{k_{{\rm DD}}}$ in a desirable manner. 
As shown in Fig.~\ref{fig:application}(b), by using a smaller $f_{k_{{\rm DD}}}$ more signal details are revealed. However, this approach also reduces the coupling to the
target spins.

By changing $f_{k_{{\rm DD}}}$ in the AXY sequences for every unit
with four composite pulses, we implement the Gaussian soft modulation
$\lambda(t)$ in a digitized manner while preserving the robustness of the sequences against experimental
control errors (see \cite{SuppMat} for details).
The resulting Gaussian AXY significantly enhances and resolves the
weak-spin signal by using the soft modulation to eliminate the strong
unwanted perturbation {[}see Fig.~\ref{fig:application}(c){]}. In addition, it removes all the side peaks around the
spin resonances, which is of great advantage when fitting dense signals
\cite{Mueller14} and avoids false identification of the signal peaks,
in particular at the presence of spurious resonances \cite{Loretz15,Haase16,Lang16}. Furthermore, it
allows robust, high-fidelity quantum gates on the desired nuclear qubits. We calculate the fidelity for the gate
$U_\text{target} = \text{exp}(-i \pi \sigma^z_0  \sigma_1^x /4) \otimes \mathbb{I}_2$ for the same parameters as the
spectrum given in Fig. 3(c), but choose $f_3$ such that the target
spin only performs a half rotation. We
include an energy shift of the strongly coupled nuclear spin
equivalent to the first example above. The fidelity is shown
in Fig.~\ref{fig:application}(d) for different values of possible pulse errors. It is always well above 99\%. 
On the contrary, the fidelities achieved by the Hartmann-Hahn or AXY protocol under the same condition
are very low (e.g., 57\% for AXY) because of the poor spin addressing [see Fig.~\ref{fig:application}(a),(b)].
Note that the enhanced spectral resolution by the soft control can be used to improve the controllability of interacting spin clusters~\cite{Zhao11Atomic,Shi14Sensing,Wang16,Wang17,Abobeih:1801.01196} and
nuclear-spin decoherence-free subspace~\cite{Perlin:1708.09414}.

\textit{Conclusions.--} We proposed the mechanism of soft quantum control which enables highly selective coupling between different on-resonance constituents
of composite quantum systems. The method introduces a time-dependent
modulation of the coupling constants in addition to the matching of
resonance frequencies. This results in an exponentially improved suppression
of off-resonant couplings. Furthermore, we establish an adiabatic average Hamiltonian
theory to describe interacting systems even under the presence of strong coupling terms to undesired parts of the Hilbert space.
We showed two direct applications of our protocol: an improved RWA and, when combined with DD techniques, the addressing of weakly coupled nuclear spins under the presence of strong perturbations, originating from impurities with close resonance frequencies. 
The method is of general applicability and can be useful for the coherent manipulations of quantum registers and spectroscopic challenges in a wide range of systems such as stoichiomeric rare earth ion systems, spin defects,
and single dopants in solids, as well as spin-boson systems.

\textit{Acknowledgements.--}
This work was supported by the ERC Synergy grant BioQ and the EU
Projects EQUAM and DIADEMS. J.~C. acknowledges Universit\"at Ulm for
a Forschungsbonus. We thank Mark Mitchison for his careful reading of the manuscript. J.F.H. and Z.-Y.W. contributed equally to this work.

\pagebreak
\clearpage
\begin{center}
\textbf{\large Supplemental Material:\\ Soft Quantum Control for Highly Selective Interactions among Joint Quantum Systems}
\end{center}
\setcounter{equation}{0} \setcounter{figure}{0} \setcounter{table}{0}
\setcounter{page}{1} \makeatletter \global\long\def\theequation{S\arabic{equation}}
 \global\long\def\thefigure{S\arabic{figure}}
 \global\long\def\bibnumfmt#1{[S#1]}
 \global\long\def\citenumfont#1{S#1}

\section{Efficient rotating-wave approximation}

\subsection{Simplified example: two coupled qubits}
\begin{figure}[t!]
\begin{centering}
\vspace{0cm}
 \hspace{-0.3cm} \includegraphics[width=0.95\columnwidth]{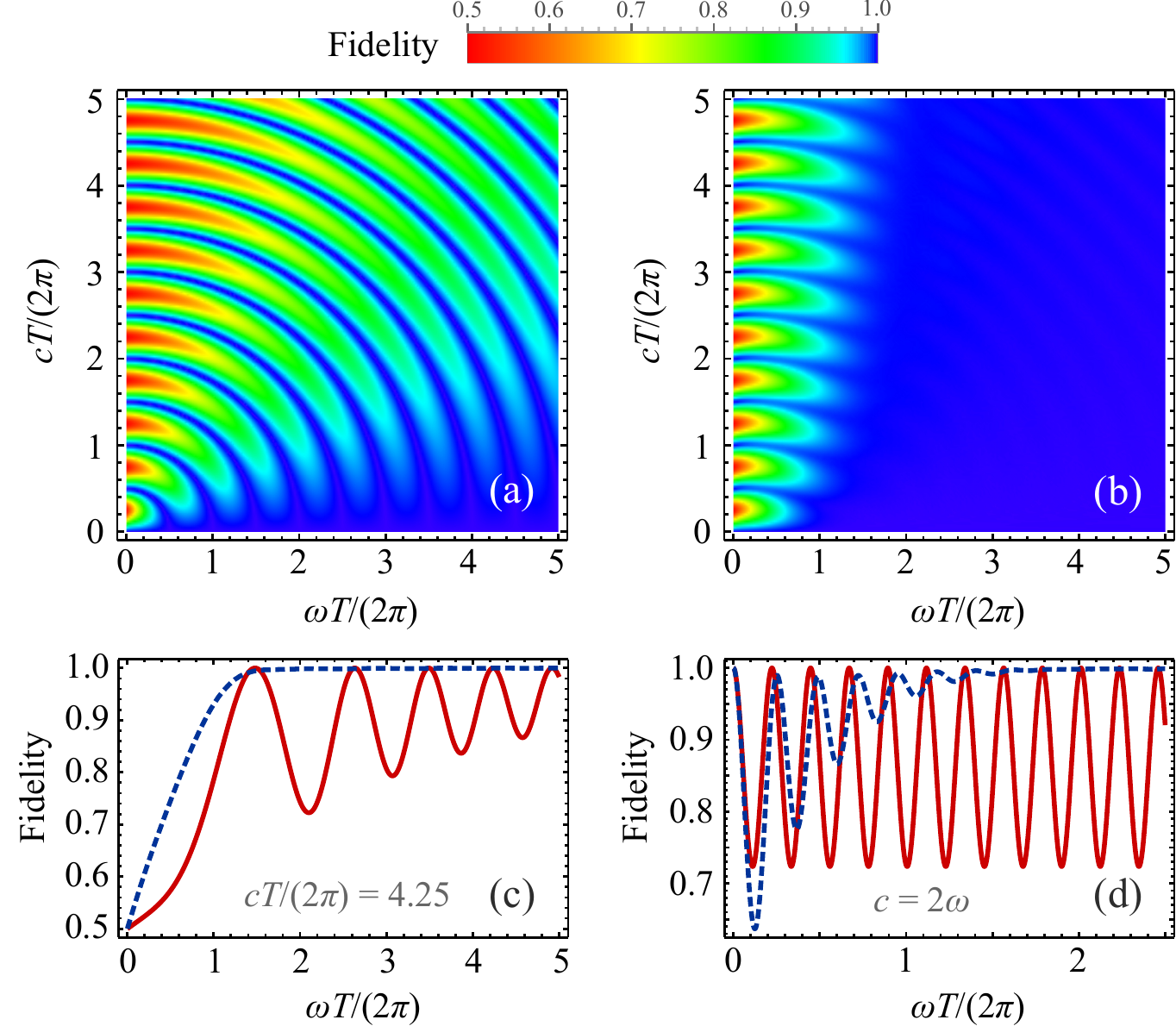} 
\par\end{centering}
\caption{(Color online) 
Fidelities to the target two-qubit dynamics. The figure is equivalent to Fig.~2 of the main text. The Gaussian shapes are realized with the parameters $\sigma=T/(4\sqrt{2})$ and $1/\lambda_{0}=\sqrt{2\pi}\sigma\text{erf}\left(\frac{T}{2\sqrt{2}\sigma}\right)$ which ensures the same target evolution as in the constant amplitude case. (a) and (b) show the constant amplitude and Gaussian modulation respectively. The curves for the Gaussian shapes [blue dashed lines in (c) and (d)] illustrate the same characteristic asymptotic behavior as for the three qubit case, while the constant amplitude undergoes significantly less modulation as only a single frequency is left in the model.}
\label{fig:smFigRWA} 
\end{figure}

To supplement the example given in the main text, we 
remove the second resource qubit of the model used in the main text (see the section titled ``Improved RWA"). 
This simplifies the illustration of the incapability of the standard approach to suppress the RWA terms efficiently.
The total Hamiltonian becomes
\begin{equation}\label{sm:eq:HtwoQubits}
H(t)=\frac{\omega}{2}\left(\sigma_{0}^{z}+\sigma_{1}^{z}\right)+c\lambda(t)\sigma_{0}^{x}\sigma_{1}^{x}. 
\end{equation}
As in the main text, we want to selectively preserve only the flip-flop interaction
between the qubits.
Explicitly, we would like to obtain the target Hamiltonian after RWA
\begin{equation}
H_{{\rm target}}=\frac{1}{2}\tilde{\omega}\, \left(\sigma_{0}^{z} +\sigma_{1}^{z}\right) + \tilde{c}\left(\sigma_{0}^{+}\sigma_{1}^{-}+{\rm H.c.}\right),
\label{sm:eq:Htarget}
\end{equation}
where the parameters take into account the energy shifts, as we explicitly show in the next section.
In Fig.~\ref{fig:smFigRWA} the equivalent of Fig.~2 of the main text is shown. While the Gaussian modulation shows a smooth transition to constantly high fidelities, the standard method achieves high fidelity only for discrete values of $\omega T$. Note that these discrete points for high fidelities in the standard method will change if there is a different number of qubits in the system (cf. Fig.~2 in the main text) and in general are hard to predict in complex systems.

\subsection{Energy shifts}
In the main text, we have shown that the soft quantum control scheme can efficiently eliminate oscillating terms while keeping desired energy-conserving
interactions. Here we illustrate the calculation of energy shifts
by considering the simple two-qubit Hamiltonian Eq.~\eqref{sm:eq:HtwoQubits} [with the target one given by Eq.~\eqref{sm:eq:Htarget}].
To simplify the calculation, we first note that the subspaces $S_{{\rm a}}=\left\{ |\uparrow\uparrow\rangle,|\downarrow\downarrow\rangle\right\} $
and $S_{{\rm b}}=\left\{ |\uparrow\downarrow\rangle,|\downarrow\uparrow\rangle\right\} $
are disconnected in $H(t)$ by denoting as $|\uparrow(\downarrow)\rangle$
the eigenstates of $\sigma_{j}^{z}$ with the eigenvalues 1 (-1).
In terms of the Pauli operators for the pseudo spin in each of the
two subspaces, the Hamiltonian $H(t)$ can be written as 
\begin{equation}
H(t)=\omega\sigma_{{\rm a}}^{z}+c\lambda(t)\sigma_{{\rm a}}^{x}+c\lambda(t)\sigma_{{\rm b}}^{x},\label{eqS:twoQubitHamiltonain}
\end{equation}
where \mbox{$\sigma_{{\rm a}}^{z}=|\uparrow\uparrow\rangle\langle\uparrow\uparrow|-|\downarrow\downarrow\rangle\langle\downarrow\downarrow|$}
and \mbox{$\sigma_{{\rm a}}^{x}=|\uparrow\uparrow\rangle\langle\downarrow\downarrow|+{\rm H.c.}$}
are the Pauli operators in the subspace $S_{{\rm a}}$ and \mbox{$\sigma_{{\rm b}}^{x}=|\uparrow\downarrow\rangle\langle\downarrow\uparrow|+{\rm H.c.}$}
is the Pauli operator in the subspace $S_{{\rm b}}$. Note that operators
for different subspaces commute. The total field strength on the pseudo
spin in the subspace $S_{{\rm a}}$ is $\sqrt{\omega^{2}+c^{2}\lambda^{2}(t)}$
and the field strength for $S_{{\rm b}}$ is $c\lambda(t)$ which can be obtained by diagonalizing~Eq.~(\ref{eqS:twoQubitHamiltonain}). Similarly,
we have 
\begin{equation}
H_{{\rm target}}=\tilde{\omega}\sigma_{{\rm a}}^{z}+c\sigma_{{\rm b}}^{x},
\end{equation}
with the corresponding field strengths of $\tilde{\omega}$ and $\tilde{c}=c$.
For adiabatic modulation on $\lambda(t)$ (including the cases for
$\lambda(t)=1$ and soft coupling), the dynamic phases driven by $H(t)$
are the same as the case for $H_{{\rm target}}$. As a consequence
we obtain the energy 
\begin{equation}
\tilde{\omega}=\frac{1}{T}\int_{-T/2}^{T/2}\sqrt{\omega^{2}+c^{2}\lambda^{2}(t)}dt \label{sm:eq:omegaTilde}
\end{equation}
by comparing the dynamic phase for the subspace $S_{{\rm a}}$. The
energy shifts $\tilde{\omega}-\omega$ include high-order effects
in the average Hamiltonian theory. The dynamic phase
for the subspace $S_{{\rm b}}$ is the average of the modulation $\frac{1}{T}\int_{-T/2}^{T/2}\lambda(t)dt=1$.

\section{Electron-nuclear systems}

\subsection{Effective Hamiltonian}

Consider the case of an NV electron spin coupled to its nearby nuclear
spins. After the secular approximation that neglects the electron-spin flip terms, 
the Hamiltonian of this system reads~\cite{SZhao11Atomic,SCasanova2015robust,SWang16} 
\begin{equation}
H_{\rm sys}=D S_{z}^{2}-\gamma_{e}B_{z}S_{z}-\sum_{j}\gamma_{j} B_{z}I_{j}^{z}+ S_{z}\sum_{j}\vec{A}_{j}\cdot\vec{I}_j+H_{\rm n,n}, \label{HNVsys}
\end{equation}
where $S_{z}$ is the spin-1 operator of the NV electron spin projected on the NV symmetry axis (i.e., the $z$ direction with the unit vector $\hat{z}$), $\vec{I}_{j}=\hat{x}I_{j}^{x}+\hat{y}I_{j}^{z}+\hat{z}I_{j}^{z}$ is the spin-$\frac{1}{2}$ operator for the $j$th nuclear spin, the zero-field splitting $D\approx 2\pi\times 2.87$GHz,  $B_{z}$ is the magnetic filed applied along the $\hat{z}$ direction, $\gamma_{e}\approx -2\pi\times2.8$ MHz/G and $\gamma_{j}$ are the gyromagnetic ratios of electron and $^{13}$C nuclear spins respectively, $\vec{A}_{j}$ describes the hyperfine interaction at the locations of the nuclear spins, and $H_{\rm n,n}$ is the nuclear-nuclear dipolar coupling.
For simplicity, the terms related to the nitrogen spin in the NV center have been neglected in the Hamiltonian $H_{\rm sys}$, because their effects can be eliminated 
by dynamical decoupling (DD)~\cite{SYang11,SSouza12,SCasanova2015robust,SWang16} and (or) by the polarization of the nitrogen spin~\cite{SEpstein2005}.
We perform simulations according to the Hamiltonian $H_{\rm sys}$. 

We select two NV electron
spin levels $0$ and $m_{{\rm s}}$ ($m_{{\rm s}}=+1$ or $-1$) to form the control qubit (with the Pauli operator $\sigma^{z}_{0}=|m_{{\rm s}}\rangle\langle m_{{\rm s}}|-|0\rangle\langle 0|$).
For the simulations of robustness, the DD pulses are applied via the microwave field 
\begin{equation}
H_{\rm ctr}= \sqrt{2}\Omega\cos(\omega_{\rm mw} t + \phi) S_{x},\label{HNVctr}
\end{equation}
where $\Omega$ is the Rabi frequency of the rectangular pulses, $\omega_{\rm mw}$ is the microwave frequency that are set to 
the frequency of the NV control qubit with a detuning $\Delta=\omega_{\rm mw}-(D-m_{{\rm s}}\gamma_{e} B_{z})$,
$\phi$ is an initial phase that control the effective directions of the DD pulses. 
In the rotating frame of the NV electron spin Hamiltonian $D S_{z}^{2}-\gamma_{e}B_{z}S_{z}$, $H_{\rm ctr}$ realizes the 
driving 
\begin{equation}
H_{\rm ctr}^{\rm qubit}=\frac{1}{2} \Omega [\sigma^{x}_{0}\cos(\phi)+\sigma^{y}_{0}\sin(\phi)], \label{Hrabi}
\end{equation} 
on the control qubit with a Rabi frequency $\Omega$. Here $\sigma^{x}_{0}=|m_{{\rm s}}\rangle\langle 0|+|0\rangle\langle m_{{\rm s}}|$.

In the following, we consider fast DD pulses that the exchange of the spin levels $0$ and $m_{{\rm s}}$  (i.e., $\sigma^{z}_{0}\rightarrow-\sigma^{z}_{0}$)
by each $\pi$ pulse can be treated as instantaneously.
Under the control of DD $\pi$ pulse sequences, the relevant Hamiltonian in the rotating frame becomes~\cite{SCasanova2015robust,SWang16}
\begin{equation}
H_{{\rm e,n}}=\frac{m_{{\rm s}}}{2}F(t)\sigma_{0}^{z}\sum_{j}\vec{I}_{j}\cdot\vec{A}_{j}-\sum_{j}\vec{I}_{j}\cdot\vec{\omega}_{j},
\end{equation}
where the vectors
\begin{equation}
\omega_{j}=\omega_{j}\hat{\omega}_{j}=\gamma_{j}B_{z}\hat{z}-\frac{1}{2}m_{{\rm s}}\vec{A}_{j}
\end{equation}
describe the nuclear precession frequencies (effective nuclear Zeeman
energies) $\omega_{j}=|\vec{\omega}_{j}|$.
The modulation function $F(t)=(-1)^{m(t)}$ when a number $m(t)$  of
$\pi$ pulses that has been applied until the time $t$.  For homonuclear spin clusters such as the case of $^{13}{\rm C}$
spins in diamond, the gyromagnetic ratios $\gamma_{j}\approx 2\pi\times1.07$ kHz/G 
are identical, but $\omega_{j}$ are shifted by the hyperfine coupling at the locations of the nuclei.
In the rotating frame with respect to $-\omega_{n}\sum_{j}\vec{I}_{j}\cdot\hat{\omega}_{j}$,
where $\omega_{n}$ is the target nuclear frequency to be addressed,
the Hamiltonian becomes 
\begin{equation}
H_{{\rm e,n}}^{\prime}=\frac{m_{{\rm s}}}{2}F(t)\sigma_{0}^{z}\sum_{j}\vec{I}_{j}\cdot\vec{A}_{j}(t)-\sum_{j}\delta_{j,n}\vec{I}_{j}\cdot\hat{\omega}_{j},
\end{equation}
where $\delta_{j,n}=\omega_{j}-\omega_{n}$ are the frequency differences
and 
\begin{equation}
\vec{A}_{j}(t)=\vec{a}_{j}^{x}\cos(\omega_{n}t)+\vec{a}_{j}^{y}\sin(\omega_{n}t)+\vec{a}_{j}^{z},
\end{equation}
with $\vec{a}_{j}^{x}\equiv\vec{A}_{j}-\vec{a}_{j}^{z}$, $\vec{a}_{j}^{y}\equiv\hat{\omega}_{j}\times\vec{A}_{j}$,
$\vec{a}_{j}^{z}\equiv a_{j}^{\parallel}\hat{\omega}_{j}$. The magnitudes
$a_{j}^{\parallel}=\vec{A}_{j}\cdot\hat{\omega}_{j}$ and $|\vec{a}_{j}^{x}|=|\vec{a}_{j}^{y}|=a_{j}^{\perp}$.
Because $\vec{a}_{j}^{\alpha}$ ($\alpha=x,y,z$) are in orthogonal
directions, we denote the spin operators $I_{j}^{\alpha}=\vec{I}_{j}\cdot\vec{a}_{j}^{\alpha}/|\vec{a}_{j}^{\alpha}|$
and $I_{j}^{\pm}=I_{j}^{x}\pm iI_{j}^{y}$. Now we can write
\begin{equation}
H_{{\rm e,n}}^{\prime}=\frac{m_{{\rm s}}}{4}F(t)\sigma_{0}^{z}\left[\left(a_{j}^{\perp}I_{j}^{+}e^{-i\omega_{n}t}+{\rm h.c.}\right)+a_{j}^{\perp}I_{j}^{z}\right]-\sum_{j}\delta_{j,n}I_{j}^{z}.
\end{equation}

We choose $F(t)$ to be symmetric and periodic $F(t+2\pi/\omega_{\text{DD}})=F(t)$, hence it can be represented by a Fourier
series, $F(t)=\sum_{k}^{\infty}f_{k}\cos(k\omega_{\text{DD}}t)$ with
the DD frequency $\omega_{\text{DD}}$ and $f_{k}=0$ for even $k$.
We tune the $k_{{\rm DD}}$-th harmonic on resonance with the target nuclear
frequency $\omega_{n}$, i.e. $k_{{\rm DD}}\omega_{{\rm DD}}=\omega_{n}$.
The nuclear spin frequencies $\omega_{j}\sim\gamma_{j}B_{z}$ can
be made significantly larger than the hyperfine coupling in experiments
by using a strong magnetic field. As a consequence, one can safely
apply RWA to remove the counter-rotating terms in $H_{{\rm e,n}}^{\prime}$,
which yields
\begin{equation}
H_{{\rm e,n}}^{\prime}=\frac{m_{{\rm s}}}{4}f_{k_{{\rm DD}}}\sigma_{0}^{z}\sum_{j}a_{j}^{\perp}I_{j}^{x}-\sum_{j}\delta_{j,n}I_{j}^{z}.
\end{equation}
Under strong magnetic fields, $\hat{\omega}_{j}\approx\hat{z}$ is
along the magnetic field direction. As a consequence, $a_{j}^{\parallel}\approx\vec{A}_{j}\cdot\hat{z}$
and $a_{j}^{\perp}$  are the parallel and perpendicular
components of  $\vec{A}_{j}$ with respect to the magnetic
field $B_{z}\hat{z}$. The effective nuclear Larmor frequencies
$\omega_{j}\approx\gamma_{j}B_{z}-\frac{1}{2}m_{{\rm s}}a_{j}^{\parallel}$
and detunings $\delta_{j,n}\approx\frac{1}{2}m_{{\rm s}}(a_{j}^{\parallel}-a_{n}^{\parallel})$
are determined by the hyperfine fields at the locations of nuclei.

For the case that $m_{{\rm s}}=-1$ and $^{13}{\rm C}$ spins in diamond
$I_{j}^{\alpha}=\frac{1}{2}\sigma_{j}^{\alpha}$, the Hamiltonian
$H_{{\rm e,n}}^{\prime}$ becomes the one of Eq.~(11) in the main text 
\begin{equation}
H=-\frac{1}{8}f_{k_{{\rm DD}}}\sigma_{0}^{z}\sum_{j>0}a_{j}^{\perp}\sigma_{j}^{x}-\sum_{j>0}\frac{1}{2}\delta_{j,n}\sigma_{j}^{z}.
\end{equation}

For CPMG or the XY family of sequences, $\omega_{\text{DD}}$ is fixed
by the application of $\pi$ pulses at times $t_{p}=\pi(p-1/2)/\omega_{\text{DD}}$,
$p=1,2,\dots$. However, a strategy like that yields constant Fourier
coefficients $f_{k}$ which are not tunable.

\subsection{AXY sequences}
The coupling coefficients $f_{k}$ can be tuned via changing the pulse intervals.
The AXY sequences employ robust composite pulses to tune $f_{k_{{\rm DD}}}$
in a desirable way~\cite{SCasanova2015robust}. As shown in Fig.~\ref{fig:properties}(a),
each of the composite pulses consists of five elementary $\pi$ pulses
in a symmetric sequence, and the relative timing of the elementary
$\pi$ pulses determines the effective values of $f_{k_{{\rm DD}}}$.
In an AXY sequence~\cite{SCasanova2015robust}, the timing of each composite pulses is 
the same such that $f_{k_{{\rm DD}}}$ has the same value though out the sequence.

\subsection{Gaussian AXY sequences}
\begin{figure}[t!]
\begin{centering}
\vspace{0cm}
 \hspace{-0.3cm} \includegraphics[width=0.95\columnwidth]{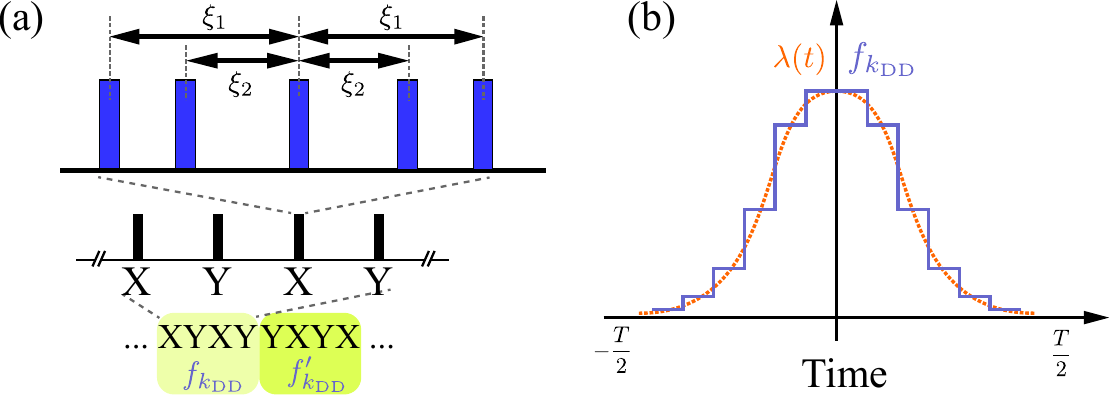} 
\par\end{centering}
\caption{(Color Online) 
Gaussian AXY sequence.
(a) The upper part shows the symmetric configuration of an AXY composite pulse,
where the relative times $\xi_{1}$ and $\xi_{2}$ among the elementary pulses determine the value of $f_{k_{\text{DD}}}$. 
The lower part shows that each four composite pulses build a robust XYXY or YXYX
block of the same $f_{k_{\text{DD}}}$. Each block in the AXY sequence employs a different
$f_{k_{\text{DD}}}$ for a soft modulation protocol in a manner that resembles the Gaussian shape as shown in (b).
In the simulation for the results in the main text, we used the Gaussian shapes with $\sigma=T/(4\sqrt{2})$. 
}
\label{fig:properties} 
\end{figure}

The ability to continuously tune $f_{k_{{\rm DD}}}$ by 
changing the relative timing of the elementary $\pi$ pulses [see Fig.~\ref{fig:properties}(a)]
allows to simulate the modulation $\lambda(t)$ for soft quantum control.
For the case that
the change of $f_{k_{{\rm DD}}}$ resembles a Gaussian shape [see  Fig.~\ref{fig:properties}(b)], we call
the sequence Gaussian AXY sequences.
In order to preserve the robustness
of the sequences~\cite{SCasanova2015robust}, we only change $f_{k_{{\rm DD}}}$ in the AXY-8
sequences for every unit with four composite pulses (XYXY or YXYX).
In this manner, the first-order control errors are cancelled completely,
while the second-order errors are small because of the symmetric XY-8 block (XYXYYXYX).

\subsection{Hartmann-Hahn resonance}
For the case of Hartmann-Hahn resonance, one can continuously drive the NV electron spin
via the control Eq.~\eqref{HNVctr} for a sensing time $T$ and scan the Rabi frequency $\Omega = \Omega_{\rm Rabi}$~\cite{SCai13,SLondon13Detecting}.
Under the continuous driving, we have the Hamiltonian for the relevant levels of the NV electron qubit,
\begin{equation}
H_{\rm H-H}= H_{\rm ctr}^{\rm qubit}-\sum_{j}\gamma_{j} B_{z}I_{j}^{z}+ m_{{\rm s}}|m_{{\rm s}}\rangle\langle m_{{\rm s}}|\sum_{j}\vec{A}_{j}\cdot\vec{I}_j,
\end{equation}
where $H_{\rm ctr}^{\rm qubit}$ is given by Eq.~\eqref{Hrabi}  and we have neglected the weak nuclear-nuclear interaction $H_{\rm n,n}$ for simplicity.
$H_{\rm H-H}$ is equivalent to 
\begin{equation}
H_{\rm H-H}^{\prime}= \frac{\Omega_{\rm Rabi}}{2}\sigma_{0}^{x} + \frac{m_{{\rm s}}}{2}\sigma_{0}^{z}\sum_{j}\vec{A}_{j}\cdot\vec{I}_j-\sum_{j}\gamma_{j} B_{z}I_{j}^{z} + \frac{1}{2}\sum_{j}\vec{A}_{j}\cdot\vec{I}_j,
\end{equation}
where for simplicity we have chosen $\phi=0$ and $\Omega = \Omega_{\rm Rabi}$ in $H_{\rm ctr}^{\rm qubit}$. One may define $\tilde{\sigma}_{0}^{z}=\sigma_{0}^{x}$
and $\tilde{\sigma}_{0}^{x}=\sigma_{0}^{z}$. By scanning the dressed-state energy $\Omega_{\rm Rabi}$ to the
the energies $\omega_{j}$ of nuclear spins, electron-nuclear flip-flop process is no longer suppressed by the energy mismatch and change the population of the electron qubit, resulting in signals in the spectrum~\cite{SCai13,SLondon13Detecting}.

\subsection{Simulation details}
In the simulations of the NV center, we use the system Hamiltonian Eq.~\eqref{HNVsys} that include 
electron-nuclear and nuclear-nuclear interactions. The effect of nuclear-nuclear interaction
is negligible because it is weak compared with the time scale of the simulations and the electron-nuclear coupling.
To take into account the realistic situations that the control has detuning and amplitude errors, we use the control Hamiltonian Eq.~\eqref{HNVctr}.
We do not consider phase error in the control Eq.~\eqref{HNVctr} because it is negligibly small in typical NV experiments that use an arbitrary waveform generator~\cite{SZopes17}. 

The NV electron spin is located
at $(0,0,0)$ while the external magnetic field of $B_{z}=400$~G is
applied along the NV symmetry axis along the direction $(1,1,1)/\sqrt{3}$.
In this basis, the two $^{13}{\rm C}$ nuclei giving the resonance
peaks in the main text are located at the diamond lattice sites $(0.26775,0.62475,0.80325)$
nm and $(0.80325,0.08925,0.08925)$ nm, 
with the corresponding $\omega_{1}=2\pi\times441.91$ kHz, $\omega_{2}=2\pi\times437.54$ kHz,
$a_{1}^{\perp}=2\pi\times16.91$ kHz, and $a_{2}^{\perp}=2\pi\times54.26$ kHz. In sensing nuclear spins, we initialize the NV electron qubit
to an eigenstate of $\sigma_{0}^{x}$ and measure the change of the population on the initial state of the NV electron qubit [Fig.~3(a)-(c) in the main text].
In the simulation for the quantum gate [Fig.~3(d) in the main text], we take the shifted energies of the nuclear spins, similarly to Eq.~(\ref{sm:eq:omegaTilde}) into account.
We would also like to point out that in the experiments on NV centers generally the fluctuation on the Rabi frequency (e.g., $2.4\times 10^{-3}$ in \cite{SCai12}) is 
much smaller than the Rabi error ($5\%$) which we used in the simulation of the main text. The detuning error $\Delta$ on the microwave field
mainly originates from the unpolarized nitrogen nuclear spin intrinsic to the NV center \cite{SLoretz15}, and by polarizing 
the nitrogen spin the detuning error becomes negligible in the AXY sequences.
Note that our simulated results are also valid when there is a large nuclear spin bath (e.g., without the use of an isotopically-purified diamond sample),
because the DD sequences can efficiently suppressed the effect of other nuclear spins that have distinct frequencies than the frequency of the target nuclear spins.
For example, using a DD sequence that has a lower spectral resolution and hence lower nuclear-spin addressability than the AXY and the soft Gaussian AXY sequences simulated in the main text, 
Ref.~\cite{SAbobeih:1801.01196} has already experimentally demonstrated selective addressing of 19 nuclear spins in a diamond with a natural abundance (1.1\%) of $^{13}$C spins.


\begin{thebibliography}{10}
\bibitem{Note:Detuning} Although the special matching condition $T\delta_{j,k}/2=n\pi$ ($n=1,2,\ldots$) for $g(\delta_{j,k})=0$ can be achieved for a single RQ with special values of the evolution time $T$, it is hard to accomplish for multiple RQs.
\end{thebibliography}
\end{document}